\documentclass[12pt]{article}
\usepackage{a4wide}
\usepackage{psfrag}
\usepackage{graphics}
\usepackage{amsfonts}
\usepackage{amstext}
\usepackage{amsmath}
\usepackage{amssymb}
\usepackage{amsthm}
\usepackage[mathscr]{eucal}
\usepackage{cite}
\def\a{\alpha}

\def\Ai{{\rm Ai}}

\def\e{\epsilon}
\def\g{\gamma}
\def\G{\mathcal{G}}
\def\i{\infty}

\def\K{\mathcal{K}}
\def\l{\lambda}

\def\o{\omega}

\def\P{\mathbb{P}}

\def\s{\sigma}

\def\t{\tau}

\begin{document}       
\title{\bf{Polynuclear growth model, GOE$^2$ \\ and \\
random matrix with deterministic source}}
\author{
\renewcommand{\thefootnote}{\arabic{footnote}}
\vspace{5mm}
T. Imamura \footnotemark[1] ~and T. Sasamoto \footnotemark[2]
}

\author{
\vspace{5mm}
T. Imamura 
{\footnote {\tt e-mail: imamura@monet.phys.s.u-tokyo.ac.jp}}
~~and
T. Sasamoto 
{\footnote {\tt e-mail: sasamoto@stat.phys.titech.ac.jp}}
 \\
{\it$^*$Department of Physics, Graduate School of Science,}\\
{\it University of Tokyo,}\\
{\it Hongo 7-3-1, Bunkyo-ku, Tokyo 113-0033, Japan}\\
\vspace{0mm}\\
{\it$^{\dag}$Department of Physics, Tokyo Institute of Technology,}\\
{\it Oh-okayama 2-12-1, Meguro-ku, Tokyo 152-8551, Japan}\\
}

\date{}
\maketitle

\begin{abstract}
We present a random matrix interpretation of the distribution 
functions which have appeared in the study of the one-dimensional 
polynuclear growth (PNG) model with external sources.
It is shown that the distribution, GOE$^2$, which is defined 
as the square of the GOE Tracy-Widom distribution,  
can be obtained as the scaled largest eigenvalue distribution 
of a special case of a random matrix model with a deterministic 
source, which have been studied in a different context previously.
Compared to the original interpretation of the GOE$^2$ as 
``the square of GOE'', ours has an advantage that 
it can also describe the transition from the GUE Tracy-Widom
distribution to the GOE$^2$.

We further demonstrate that our random matrix interpretation 
can be obtained naturally by noting the similarity of the 
topology between a certain non-colliding Brownian motion model 
and the multi-layer PNG model with an external source.  
This provides us with a multi-matrix model interpretation of 
the multi-point height distributions of the PNG model with 
an external source.
\vspace{3mm}

\noindent
[Keywords: KPZ universality, random matrices, 
polynuclear growth model]

\end{abstract}
\newpage
\section{Introduction}
The universality of eigenvalue correlations in random matrix 
theory(RMT) has been arising in various fields in physics. 
The most recent appearance is in the one-dimensional 
Kardar-Parisi-Zhang(KPZ) universality class~\cite{KPZ1986}
which is a fundamental universality 
in non-equilibrium statistical mechanics. 
It has been found that the fluctuation of 
certain physical quantities in the KPZ universality class 
is equivalent to that of the largest eigenvalue in RMT.
This fact was first pointed out in the polynuclear growth(PNG) 
model~\cite{PS2000b} and the totally asymmetric simple exclusion
process(TASEP)~\cite{Jo2000} based on the Baik-Deift-Johansson
theorem in random permutations~\cite{BDJ1999}.

The PNG model is a stochastic surface growth model and
is the most well-understood model in the relation with RMT. 
Based on the results on the symmetrized random 
permutations~\cite{BR2001a,BR2001b,BR2001c}, 
the properties of the height fluctuation of the PNG model
have been analyzed for various boundary conditions.  
In~\cite{PS2000b} the height fluctuation of the PNG droplet
in an infinite space was analyzed. It turned out that
the height fluctuation is described by the GUE Tracy-Widom 
distribution~\cite{TW1994}.
On the other hand, in case of the growth on half line, the height
fluctuation at the boundary is described by the GOE and GSE 
Tracy-Widom distribution~\cite{TW1996} 
according to the strength of an external 
source at the boundary~\cite{PS2000a}. 
Furthermore the equal time correlation of the height fluctuation
at distinct points has also been analyzed for the PNG model. 
It has been found that it is equivalent to the process of the largest 
eigenvalue in the Dyson's Brownian motion model between GUEs in the 
case of the growth on infinite line~\cite{PS2002b,Johansson2002p} 
while in half infinite case, it 
corresponds to the process in the model which describes the 
transition between GOE/GSE and GUE~\cite{SI2004}.

In addition to the Tracy-Widom distributions for the 
fundamental three classes in RMT, GOE, GUE and GSE,  
two distribution functions, which had not been discussed in the study 
of RMT, arose in the description of the height fluctuation of the PNG 
droplet on infinite line with external sources at boundaries
~\cite{PS2000a,BR2000}. 
These are called $F_0$ and GOE$^2$. 
These functions also appear in the study of current fluctuations in 
TASEP~\cite{PS2002a, NS2004p}. 
There are also distribution functions which describe the 
transitions between two functions mentioned above. For instance,
there is a distribution function, depending on a parameter $\o$,
which approach the GUE Tracy-Widom distribution as $\o\to \i$, 
tends to the Gaussian as $\o\to -\i$ and becomes GOE$^2$ when $\o=0$.
We have not yet known what kind of 
random matrix ensemble corresponded to $F_0$. 
The GOE$^2$, on the other hand, can be understood in terms of the language 
of RMT. It is written as the square of the GOE Tracy-Widom distribution
and can be interpreted as the distribution function of the larger of 
the largest eigenvalues of two independent GOEs. 
Its Fredholm determinant representation was obtained in 
~\cite{Forrester2000p}.
There, it was also shown that the scaling limit of certain 
eigenvalue statistics of two independent GOEs is described by 
its kernel. 
        
In this paper, we present another point of view of the GOE$^2$ 
as the scaled largest eigenvalue distribution of a random matrix 
with a deterministic source. 
The advantage of this ensemble is that it can describe the 
GUE-GOE$^2$ transition in the PNG model, which is induced by 
the change of the external source while this transition 
cannot be obtained by the measure of ``the square of GOE''.

{From the point of view of the PNG model with external sources,
our viewpoint is more natural than that of ``the square of GOE''. 
We turn our attention on the topology and 
the structure of the determinantal measure in the multi-layer 
version of the PNG model with an external source.
Our random matrix ensemble is obtained by noticing that a certain 
type of the non-colliding Brownian motion has a similar topology 
and the structure of the measure.}

The random matrix obtained in this paper is a special case of the
random matrix with a deterministic source, which has been 
studied in~\cite{BH1,BH2,BH3,BH4,BH5,BH6,BH7, PZ1,PZ2}. 
In a special case, which is different from ours, its eigenvalue 
statistics has an interesting property that the level spacing 
distribution becomes an intermediate one between the Poisson and 
the Wigner-Dyson statistics. This property connects it with the 
fields in physics such as the quantum chaos, the metal-insulator 
transition in disordered systems and so on.
Recently the universality of such eigenvalue statistics in this model 
have been analyzed by the Riemann-Hilbert technique~\cite{BK1,BK2,BK3,BK4}.

This paper is arranged as follows. In the next section we summarize
the results on the height distribution in the PNG model with an external 
source. Then we introduce a random matrix with a deterministic source 
and compare its largest eigenvalue statistics with the height
fluctuation of the PNG model. In section \ref{ffermion} 
we consider a reason why such 
a random matrix model appears in the context of the PNG model with an 
external source. We notice that the multi-layer version of the PNG model 
can be regarded as a non-colliding many body random walks (vicious walk)
and hence inherits a free fermionic picture.
By considering a certain vicious walk which is expected to have a
similar property to the multi-layer PNG model, we see there appears naturally 
a Dyson's Brownian motion model (a multi-matrix model) involving an 
external source in the initial condition. Finally
we show the equal-time correlation of the PNG model
with an external source is equivalent to the dynamical correlations 
of the largest eigenvalue in the Dyson's Brownian motion's model.
The last section is devoted to the conclusion.

\section{PNG model and random matrix 
with deterministic source}
\label{PNGRMT}
\subsection{PNG model with external source}
First of all we give the definition of the one-dimensional 
discrete PNG model~\cite{Johansson2002p}.
Let $r\in \{\cdots, -1, 0, 1,\cdots \}$ and 
$t\in\{1, 2,\cdots\}$ be the space and time coordinate respectively.
The model is a stochastic surface growth model and consists of the 
rules (a)$\sim$(c).  Fig.1 illustrates these rules. 
\begin{itemize}
\item[(a)]  At time $t$, a nuclear is generated randomly at 
$r=-t+1,-t+3,\cdots,t-3,t-1$. The height $k$ ($=0,1,2,\cdots$)
of each nuclear is independent of $r$ and obeys the geometric 
distribution with some parameter which is fixed shortly.
\end{itemize}
This is called nucleation. While a nucleation is stochastic, 
a growth after the nucleation given by the rules (b) and (c) is 
deterministic .
\begin{itemize}
\item[(b)] Each nuclear grows laterally with one step toward right and left 
at each time step and then forms a step.
\item[(c)] When a step crashes against another step, 
the height of the crashing point is that of the higher one.
\end{itemize}
These rules (a)$\sim$ (c) can be formulated by a single equation,
\begin{equation}
  h(r,t+1)=\text{max}\left[h(r-1,t),h(r,t), h(r+1,t)\right]+\o(r,t+1).
\label{rule1}
\end{equation}
Here $h(r,t)$ is the height of the surface at time $t$ and at position
$r$, and $\o(r,t)$ means a height of a nucleation which is a random variable. 

In this paper we set the random variable $\o(r,t)$ as follows.
\begin{itemize}
\item When $t-|r|>0$ and $t-r$ is odd,
  \begin{equation*}
\quad\P[\o(r,t)=k]=
\begin{cases}
  (1-\a\sqrt{q})\left(\a\sqrt{q}\right)^k & r=-t+1,\\
  (1-q)q^k & \text{otherwise},
\end{cases}
\end{equation*}
\item Otherwise
  \begin{equation}
  \o(r,t)=0.
  \label{rule2}
  \end{equation}
\end{itemize}
Because of the condition that the nucleation occurs only when $t-|r|>0$,
the surface grows into a droplet shape.
The condition that $t-r$ is odd is imposed only for a technical 
reason and is irrelevant to the physical properties. 
Basically $\o(r,t)$ is a geometric random variable with a parameter 
$q$. Only on the left edge, the parameter is
$\a\sqrt{q}(\sqrt{q}\le\a<\frac{1}{\sqrt{q}})$. Thus the parameter $\a$
adjusts the boundary condition as an external source.    
This situation representing~\eqref{rule2} is
depicted in Fig.2. 

In the setting of~\eqref{rule1} and~\eqref{rule2}, we focus on the 
distribution of the scaled height at the origin $H_N$ defined as
\begin{equation}
\label{HN0}
  H_N=\frac{h(r=0,t=2N)-aN}{dN^{\frac13}},
\end{equation}
where $a=\frac{2\sqrt{q}}{1-\sqrt{q}}, d=\frac{(1+\sqrt{q})^{\frac13}
q^{\frac16}}{1-\sqrt{q}}$. The limiting distribution has been 
obtained in ~\cite{PS2000b,PS2000a, BR2000}. 
It depends dramatically on the value of $\a$. The result is
\begin{equation}
 \lim_{N \to \i} \P[H_N\le s] =
 \begin{cases}
   F_2(s),&\text{for}\quad\sqrt{q}\le\a<1,\\
   F_1(s)^2,&\text{for}\quad\a=1, \\
     0,   &  \text{for}\quad 1<\a<\frac{1}{\sqrt{q}},
 \end{cases}
\label{PNGresult}
\end{equation}
where $F_2(s)$ and $F_1(s)$ are the GUE and GOE Tracy-Widom distribution
respectively~\cite{TW1994, TW1996}. 
$F_2(s)$ is defined as follows. Let $\l_1$ be the largest
eigenvalue in $N\times N$ GUE random matrix with the measure,
\begin{equation}
  e^{-\text{tr} {M_2}^2}dM_2,
\end{equation}
where $M_2$ is an $N\times N$ Hermitian matrix. 
Taking the edge scaling 
for $\l_1$,
\begin{equation}
\label{escaling}
  \l_1=\sqrt{2N}+\frac{X_1}{\sqrt{2}N^{\frac16}},
\end{equation}
we define $F_2(s)$ as the limiting distribution of $X_1$,
\begin{equation}
  F_2(s)\equiv\lim_{N\rightarrow\i}\P[X_1\le s].
\end{equation}
$F_1(s)$ is also defined in the same way as above for $N\times N$
GOE random matrix with the measure,
\begin{equation}
   e^{-\frac{\text{tr} {M_1}^2}{2}}dM_1,
\end{equation}
where $M_1$ is an $N\times N$ real symmetric matrix.
Hence the distribution $F_1(s)^2$, can be interpreted as 
the distribution of the largest eigenvalue 
in the superimposition of eigenvalues of two independent GOE random matrices. 
In this sense the distribution 
defined by $F_1(s)^2$ is denoted as GOE$^2$. 

Both $F_2(s)$ and $F_1(s)^2$ can be represented as the Fredholm
determinant,
\begin{align}
  F_2(s)=\det\left[1+\K_2 g\right],\quad
  F_1(s)^2=\det\left[1+\K_{12} g\right],
\end{align}
where $g(x)=-\chi_{(s,\i)}(x)$.
The definition of the Fredholm determinant with kernel $\K(x,y)$ is 
\begin{equation}
\det[1+\K g]=\sum_{k=0}^{\i}\frac{1}{k!}
\int_{-\i}^{\i}\cdots\int_{-\i}^{\i} dx_1 \cdots dx_k 
g(x_1)\cdots g(x_k) \det[\K(x_i,x_j)]_{i,j=1}^{k}.
\label{Fred}
\end{equation}
The kernels $\K_2(x,y)$ and $\K_{12}(x,y)$ have the following form,
\begin{align}
\K_2(x,y)&=\int_0^{\i}d\s\Ai (x+\s)\Ai (y+\s),\notag \\
\K_{12}(x,y)&=\K_2(x,y)+\Ai(x)\int_0^{\i}d\s\Ai(y-\s). 
\end{align}
Here $\K_2(x,y)$ is called the Airy 
kernel~\cite{Forrester1993} and 
$\K_{12}(x,y)$ is obtained in~\cite{Forrester2000p}.

Thus when $\a\le 1$, the result~\eqref{PNGresult} can be rewritten as 
\begin{equation}
  \lim_{N \to \i} \P[H_N\le s] =\det[1+\K g],
\end{equation}
where 
\begin{equation}
  \K(x,y)=
\begin{cases}
\K_2(x,y)=\int_0^{\i}d\s\Ai (x+\s)\Ai (y+\s), & \text{for } \sqrt{q}\le\a<1, \\
\K_{12}(x,y)=\K_2(x,y)+\Ai(x)\int_0^{\i}d\s\Ai(y-\s),&\text{for\quad} \a=1. 
\end{cases}
\label{extker}
\end{equation}

When $1<\a<\frac1{\sqrt{q}}$ , the result~\eqref{PNGresult} 
that the limiting distribution vanishes
implies that we have to define another scaled height,
\begin{equation}
  H_{N}^{(G)}=\frac{h(r=0,t=2N)-a_G N}{d_GN^{\frac12}},
\end{equation}
where $a_G=\frac{q^{\frac12}(1-2\a q^{\frac12}+\a^2)}
{(\a-q^{\frac12})(1-\a q^{\frac12})}$ and
$d_G=\frac{\a^{\frac12} q^{\frac14}(1-q)^{\frac12}(\a^{2}-1)^{\frac12}}
{(1-\a q^{\frac12})(\a-q^{\frac12})}$.
Then the limiting distribution becomes the error function, 
\begin{equation}
   \lim_{N \to \i} \P[H_N^{(G)}\le s] =\frac{1}{\sqrt{2\pi}}
\int_{-\i}^{s}d\xi e^{-\frac{\xi^2}{2}},
~~~~~~~
\text{for}~~
1<\a<\frac1{\sqrt{q}}.
\label{gauss}
\end{equation}

\subsection{Random matrix with deterministic source}
\label{rmds}
When we look at the above results,
we notice a fact that the distribution at $\a=1$, which we call 
GOE$^2$, appears in 
the intertwining point between Gaussian statistics and RMT statistics. 
First, let us consider the distribution of the height at the origin 
when $\a$ is extremely large. In this case we can easily 
expect that it is the Gaussian since the height at the origin is 
virtually determined by nucleations at the only one point, 
the left edge due to the rule (b). 
Equation \eqref{gauss} confirms this discussion.
When $\a=0$, on the other hand, it is described by the 
GUE Tracy-Widom 
distribution as discussed above.
The GOE$^2$ arises in the situation where these two statistics compete i.e.
when $\a=1$.

Considering this property, we introduce a simple random matrix model
which is expected to describe the GUE random matrix--Gaussian transition
of the largest eigenvalue. We consider a random matrix $H_0$ defined as 
\begin{equation}
  \label{eq:1}
H_0=H+V,
\end{equation}
where $H$ is the GUE random matrix and $V$ is the deterministic 
diagonal matrices, $V=\text{diag}(\e_j)~~(j=1, \cdots, N).$
The measure can be defined as
\begin{equation}
  \label{eq:3}
  e^{-\text{tr}H^2}dH_0.
\end{equation}
Focusing on the case where
\begin{equation}
  \label{eq:4}
  \e_1=\e, \e_2=\cdots=\e_N=0,
\end{equation}
we can expect that when $\e$ is sufficiently large 
the largest eigenvalue is isolated from others by the
effect of $\e$ and then the fluctuation becomes the Gaussian while
this becomes the GUE Tracy-Widom distribution when $\e$ is
sufficiently small.

The eigenvalue statistics of~\eqref{eq:1} and~\eqref{eq:3} is
analyzed 
in~\cite{BH1,BH2,
BH3,BH4,BH5,BH6,BH7,PZ1,PZ2}. 
For the distribution of the largest eigenvalue $\l_1$, their
results lead to the Fredholm determinant expression,
\begin{equation}
  \label{eq:5}
\P[\l_1\leq s]=\det[1+K_Ng]. 
\end{equation}
Here the kernel $K_N(x,y)$ can be expressed as a double integral,
\begin{equation}
  \label{eq:6}
  K_N(x,y)=-\int_\Gamma\frac{dv}{2\pi i}\int \frac{du}{2\pi}
\prod_{j=1}^{N}\left[\frac{-\e_j-\frac{iu}{2}}{v-\e_j}\right]
\frac{e^{-\frac{u^2}{4}-v^2+iu x+2vy}}{v+i\frac{u}{2}},
\end{equation}
where $\Gamma$ denotes a contour enclosing $\{\e_j\}_{j=1,\cdots,N}$
anticlockwise and the integration wrt $u$ is taken from $-\i$ to $\i$ and 
not to cross $\Gamma$. 
   
In this formula with the condition~\eqref{eq:4}, 
we consider the following scaling limit, 
\begin{equation}
  \label{eq:7}
  x=\sqrt{2N}+\frac{X}{\sqrt{2}N^{\frac16}},
  \quad 
  y=\sqrt{2N}+\frac{Y}{\sqrt{2}N^{\frac16}},
  \quad 
  \e=\Lambda\sqrt{\frac{N}{2}},
\end{equation}
with $\Lambda(\leq 1)$ fixed.
This corresponds to the edge scaling limit~\eqref{escaling} 
of the eigenvalues.
Applying the saddle 
point method, the limiting kernel can be calculated.
The result depends on the value of $\Lambda$,
\begin{equation}
\label{8}
  K_N(x,y)\rightharpoonup
\begin{cases}
\K_2(X,Y)=\int_0^{\i}d\s\Ai (X+\s)\Ai (Y+\s), & \text{for } 
\Lambda <1,\\
\K_{12}(X,Y)=\K_2(X,Y)+\Ai(X)\int_0^{\i}d\s\Ai(Y-\s),& \text{for } 
\Lambda =1. 
\end{cases}
\end{equation}
Note that the kernel~\eqref{8} is exactly the same as that in the PNG 
model with an external source~\eqref{extker} and 
$\Lambda$, the parameter of the deterministic source in RMT 
corresponds to $\a$, the parameter of the external source in PNG model. 
Hence we find that 
the height fluctuation in the PNG model with an external source
shares the limiting distribution with the largest eigenvalue
fluctuation of the random matrix with deterministic source 
defined in~\eqref{eq:1}--~\eqref{eq:4}.
The same distributions appear in the distribution of the largest 
eigenvalue of the non-null complex sample covariance 
matrices~\cite{BBAP2004p}.

When $\Lambda>1$ in~\eqref{eq:7}, we change the edge 
scaling~\eqref{eq:7} to
\begin{equation}
\label{gscaling}
 x=A_G N^{\frac12}+B_G X,
 \quad
 y=A_G N^{\frac12}+B_G Y, 
\end{equation}
where $A_G=\frac{1}{\sqrt{2}}\left(\Lambda+\frac{1}{\Lambda}\right)$ and
$B_G=\sqrt{\frac{\Lambda^2-1}{2\Lambda^2}}$.
The kernel is asymptotically 
\begin{equation}
  K_N(x,y)\sim  \K_G(X,Y)=
\frac{e^{\sqrt{2N}\Lambda(y-x)}}{\sqrt{2\pi} B_G}e^{^\frac{-X^2}{2}},
~~~~\text{for}~~\Lambda>1.
\label{gauss2}
\end{equation}
Thus we get
\begin{align}
  \lim_{N\rightarrow\i}\P\left[\frac{\l_1-A_G\sqrt{N}}{B_G}\le 
s\right]
&=\frac{1}{\sqrt{2\pi}}
\int_{-\i}^{s}d\xi e^{-\frac{\xi^2}{2}}.
\end{align}
This result corresponds to~\eqref{gauss}.
Note that the factor $e^{\sqrt{2N}\Lambda(y-x)}$ in~\eqref{gauss2} 
does not contribute to the Fredholm determinant.
 
At last we give a proof of \eqref{gauss2}. 
The main result in this section~\eqref{8} will be derived in
section \ref{12transition} under more general situation.

Rescaling the variables in~\eqref{eq:6} under the 
condition~\eqref{eq:4} such that, 
\begin{equation}
 v=\sqrt{\frac{N}{2}}z,\quad -\frac{i u}{2}=\sqrt{\frac{N}{2}}w,
 \quad \e=\Lambda\sqrt{\frac{N}{2}},    
\end{equation}
one finds
\begin{align}
 \quad {K}_N(x,y)
 =
 \frac{\sqrt{2N}}{(2\pi i)^2}
 \int_{\Gamma'}dz\int_{-i\i}^{i\i}dw\frac{z}{w}
 \frac{w -\Lambda}{z-\Lambda} 
 \frac{1}{w-z}
\frac{e^{\left(\sqrt{2N}y-\sqrt{2}A_GN\right)z}}
{e^{\left(\sqrt{2N}x-\sqrt{2}A_GN\right)w}}
 e^{N\left\{f_G(w)-f_G(z)\right\}}, 
\end{align}
where the contour $\Gamma'$ encloses the origin and $\Lambda$ anticlockwise and
\begin{equation}
  f_G(w)=\frac{w^2}{2}-\sqrt{2}A_G w+\ln w.
\label{defg}
\end{equation}
$K_N(x,y)$ can be divided into two parts,
\begin{equation}
  K_N(x,y)=K_N^{(1)}(x,y)+K_N^{(2)}(x,y).
\label{K1K2}
\end{equation}
Here $K_N^{(1)}$ and $K_N^{(2)}$ correspond to a contour integral 
of $z$ around 
$\Lambda$ and the origin respectively and have the forms, 
\begin{align}
  K_N^{(1)}(x,y)&\equiv \frac{\sqrt{2N}\Lambda}{2\pi i}
 \frac{e^{\left(\sqrt{2N}y-\sqrt{2}A_GN\right)\Lambda}}{e^{Nf_G(\Lambda)}}
  e^{-Nf_G(\Lambda)}\int_{-i\i}^{i\i}\frac{dw}{w} \frac{e^{Nf_G(w)}}
 {e^{\left(\sqrt{2N}x-\sqrt{2}A_GN\right)w}},
\label{KN1}\\
  K_N^{(2)}(x,y)&\equiv
 \frac{\sqrt{2N}}{(2\pi i)^2}
 \oint dz\int_{-i\i}^{i\i}dw\frac{z}{w}
 \frac{w -\Lambda}{z-\Lambda} 
 \frac{1}{w-z}
\frac{e^{\left(\sqrt{2N}y-\sqrt{2}A_GN\right)z}}
{e^{\left(\sqrt{2N}x-\sqrt{2}A_GN\right)w}}
 e^{N\left\{f_G(w)-f_G(z)\right\}},  
\end{align}
where $\oint$ means the contour enclosing the origin anticlockwise.
We analyze the asymptotic form of each $K_N^{(i)},~~(i=1,2)$ by the saddle
point method under the scaling~\eqref{gscaling}.
 
At first we consider $K_N^{(1)}(x,y)$. 
We deform the path of integration of $w$ to the one which 
passes a critical point $w_c=\Lambda$ of $f_G(w)$. 
We set
\begin{equation}
\label{gwchange}
  w=\Lambda+\frac{iw_1}{\sqrt{2N}B_G} 
\end{equation}
and substitute this into the integrand in~\eqref{KN1} to get 
\begin{align}
  &Nf_G(w)\sim Nf_G(\Lambda)-\frac{w_1^2}{2},\notag\\
  &e^{\left(\sqrt{2N}x-\sqrt{2}A_GN\right)w}\sim
  e^{\left(\sqrt{2N}x-\sqrt{2}A_GN\right)\Lambda}e^{iw_1X}.
\label{NGw}
\end{align}
Thus one finds
\begin{equation}
 K_N^{(1)}\sim e^{\sqrt{2N}\Lambda(y-x)}
 \frac{e^{-\frac{X^2}{2}}}{\sqrt{2\pi}B_G}.
\label{gK1}
\end{equation}

For $K_N^{(2)}(x,y)$,  we deform the path of the variable $z$ in a way that
it crosses another critical point 
$z_c=\frac{1}{\Lambda}$ of $f_G(z)$
while as for $w$, we again use~\eqref{gwchange}.
Hence $z$ is transformed to 
\begin{equation}
z=\frac{1}{\Lambda}+\frac{iw_2}{\sqrt{2N}B_G},  
\end{equation}
and then we get
\begin{align}
  &-Nf_G(z)\sim -Nf_G\left(\frac1{\Lambda}\right)-\frac{w_1^2}{2},\notag\\
  &e^{\left(\sqrt{2N}y-\sqrt{2}A_GN\right)z}\sim
  e^{\left(\sqrt{2N}y-\sqrt{2}A_GN\right)\frac1{\Lambda}}e^{iw_2Y}.
\label{NGz}
\end{align}
Due to~\eqref{NGw} and~\eqref{NGz}, one finds 
\begin{align}
  {K_N^{(2)}(x,y)}
&\sim e^{N\left\{f_G(\Lambda)-f_G\left(\frac1{\Lambda}\right)\right\}}
 e^{\mathcal{O}(N^{\frac12})}\notag\\
&=e^{-\frac{N}{2}\left(\Lambda^2-\frac{1}{\Lambda^2}-4\ln\Lambda\right)} 
  e^{\mathcal{O}(N^{\frac12})}\rightarrow 0,
\label{gK2}
\end{align}
where we use $\Lambda>1$.

Thus from~\eqref{K1K2},~\eqref{gK1} and~\eqref{gK2},
we finally get~\eqref{gauss2}.

\setcounter{equation}{0}
\section{Free fermionic picture}
\label{ffermion}
\subsection{Multi-layer PNG}
In the above section we discussed the limiting distribution of the 
scaled height at the origin in the PNG model and its dependence
on the external source. In the next step, we would like to consider
the multi-point equal time joint distributions. 
To do this, we introduce the multi-layer PNG model from the time evolution
of the PNG model~\cite{PS2002b,Johansson2002p}. 

The rule of the multi-layer PNG model is as follows.
When two steps collide, the lower step is absorbed by the higher one
due to the rule (c).  We recover this absorbed height as the nucleation
in the layer below. This situation and a typical example of 
the multi-layer PNG model are illustrated in Fig.3.  
Note that the first layer of the multi-layer PNG model represents 
the shape of the PNG droplet while other layers record the 
time evolution of the growth. 

If we treat the $i$-th layer of the multi-layer PNG model as the $i$-th 
random walker, the multi-layer PNG model can be regarded as a non-colliding 
random walks called the vicious walk. 
Thus the multi-point equal time correlation of the PNG model, which
is our target, corresponds to the dynamical correlation of the first 
walker in the non-colliding random walks. Note that we treat the space axis
in the PNG model as the time axis in the point of view of the vicious walk. 
The advantage of the mapping is that as we will see below, 
the measure can be described in the form of products of 
determinants. This determinantal structure is associated with
the structure of wavefunction in the $N$-body free fermions, 
i.e. the Slater determinant.

In~\cite{IS2004p}, we could obtain the multi-point equal time joint 
distributions in the PNG model having external sources at both edges 
by introducing the multi-layer PNG model. The result specialized 
to the case considered here is the following.  

We define the scaled height $H_N(\t)$ near the origin as
\begin{equation}
\label{scaledH}
 H_N(\t) 
 = 
 \frac{h(2cN^{\frac23}\t,t=2N)-a N}{d N^{\frac13}} +\tau^2,
\end{equation}
where $a$ and $d$ are already given below \eqref{HN0} and 
$c=\frac{(1+\sqrt{q})^{\frac23}}{q^{\frac16}}$.
We set the parameter of the external source as
\begin{equation}
\label{g-scale}
 \a = 1-\frac{\o}{dN^{\frac13}}.
\end{equation} 
The equal time multi-point distribution function near the origin 
is described by the Fredholm determinant,
\begin{align}
\label{detKlim_o0}
 &\quad\lim_{N\to\i} \P[H_N(\t_1) \leq s_1, 
   \cdots , H_N(\t_m) \leq s_m] \notag\\
 &=\det\left[1+\K \G\right],\notag\\
 &\equiv \sum_{k=0}^{\i} \frac{1}{k!} 
 \sum_{n_1=1}^m \int_{-\i}^{\i} d\xi_1 \cdots 
 \sum_{n_k=1}^m \int_{-\i}^{\i} d\xi_k 
 ~\G(\t_{n_1},\xi_1) \cdots \G(\t_{n_k},\xi_k)  
 \det\left[\K(\t_{n_l},\xi_l; \t_{n_{l'}},\xi_{l'})\right]_{l,l'=1}^k,\notag\\
\end{align}
where  $\G(\t_j,\xi)=-\chi_{(s_j,\i)}(\xi)$ ($j=1,2,\cdots,m$).
The kernel is 
\begin{equation} 
 \K(\t_1,\xi_1;\t_2,\xi_2) 
 =\K_2^{\text{ext}}(\t_1,\xi_1;\t_2,\xi_2)
  +\Ai(\xi_1) \int_0^{\i} d\l e^{-(\o+\t_2)\l}\Ai(\xi_2-\l), 
   \label{Kernel12}
\end{equation}
where
\begin{equation}
 \label{K2def}
 \K_2^{\text{ext}}(\t_1,\xi_1;\t_2,\xi_2)
 =
 \begin{cases}
  \int_0^{\i} d\l e^{-\l(\t_1-\t_2)} \Ai(\xi_1+\l) \Ai(\xi_2+\l), 
  & \t_1 \geq \t_2 ,\\
  -\int_{-\i}^0 d\l e^{-\l(\t_1-\t_2)} \Ai(\xi_1+\l) \Ai(\xi_2+\l), 
  & \t_1 < \t_2 .
 \end{cases}
\end{equation}
In the case where all $\t_i=0$, another representation of this distribution
has been obtained via Riemann-Hilbert method
by Baik and Rains~\cite{BR2000}. 

The kernel $\K_2{^\text{ext}}$ is called the extended 
Airy kernel~\cite{Mac1994} and the 
process described by this Fredholm determinant 
is called the Airy process~\cite{PS2002b,Johansson2002p}. It describes the 
process of the largest eigenvalue in the Dyson's Brownian motion
model in Hermitian matrices. It also describes the correlation function 
of the PNG model without an external source. 

Looking at the correlation
function~\eqref{detKlim_o0} at one point case ($m=1$), we can
find that it expresses the GOE$^2$--GUE transition induced
by both parameters of the external source $\o$ and a position
$\t$. Setting 
$\o=\t_1=0$ in~\eqref{Kernel12} in the one point case, we get $\K=\K_{12}$ 
which is the kernel of the edge scaling in
GOE$^2$ ensemble while in the case where $\o\rightarrow\i$ or 
$\t\rightarrow\i$ one finds $\K\sim\K_2$ and the GUE case is recovered.

\subsection{Non-colliding Brownian motion and 
multi matrix model}
In subsection \ref{rmds} the random matrix with a deterministic 
source, which corresponds to the PNG model with an external source 
was introduced from an intuitive argument. 
Here we show that the measure of the random matrix with a deterministic 
source is obtained naturally by considering
the measure of the multilayer version of the 
PNG model with an external source. The analysis not
only reproduces the result in subsection \ref{rmds} but also provides 
a multi-matrix model which has the kernel~\eqref{Kernel12} in
the edge scaling limit.

As was shown in the above subsection, 
it is found through the multi-layer PNG model that
the structure of the vicious walk
(free-fermionic structure) is hidden in the measure of the PNG model. 
Particularly we focus on the two properties in the measure of the
vicious walk. The first property is the determinantal structure
which the vicious walk and RMT share. The second one is the topology   
of the vicious walk. In our case, Fig.(3-b), the top layer, which 
corresponds to the top walker in the point of view of the vicious walk,
seem to start from the point far from the other walkers due to 
the effect of the external source.
In the study of vicious walks, it has been known that in some cases
the measure can be represented as that of a multi-matrix model in an 
appropriate scaling limit 
and its topology determines the universality classes of the measure 
in RMT~\cite{KT1,KT2}.

Keeping in mind the above two properties,
we begin our discussion with writing down the measure of 
non-colliding Brownian motion which can be regarded as the 
continuous limit of the vicious walk.
Let $x_r^{(j)}(r=1, \cdots, N, j=0,\cdots ,M+1)$ be the position of
$r$-th walker from top at the time labeled by $j$. By the 
Karlin-McGregor theorem, the measure that the walkers 
have a configuration \{$x_r^{(j)}$\} is obtained as 
a product of the determinants of propagaotrs of the one-body 
Brownian motion,
\begin{equation}
  \label{31-1}
\frac{1}{Z}\prod_{j=0}^{M}
\det\left[
\psi_{j,j+1}(x_r^{(j)},x_s^{(j+1)})
\right]_{r,s=1}^{N}.
\end{equation}
Here $Z$ is the normalization constant and 
$\psi_{j,j+1}(x_r^{(j)},x_s^{(j+1)})$ is the propagator
of the one body Brownian motion from $x_r^{(j)}$ to $x_s^{(j+1)}$
with interval $T_j$,
\begin{equation}
  \label{eq:31-2}
 \psi_{j,j+1}(x_r^{(j)},x_s^{(j+1)})
=\frac1{\sqrt{2\pi T_j}}\exp\left\{
-(x_s^{(j+1)}-x_r^{(j)})^2/2T_j
\right\}.
\end{equation}

Then we take the following scaling limit,
\begin{equation}
  T_j =r_jT~~(j=0,1,2,\cdots, M),\quad
  x_r^{(j)}=
  \begin{cases}
    \frac{\sqrt{T} \e_r}{2}, & j=0,\\
    \sqrt{s_j T}\l_r^{(j)},&j=1,2,\cdots, M,\\
    b_r,& j=M+1.
  \end{cases}
\label{scaling}
\end{equation}
Note that if we take the parameter of the initial condition to be
$\e_1=\e, \e_2=\cdots=\e_N=0$, the topology of the configuration
becomes similar to that in the multi-layer PNG model with an external source.
(See Fig.4.)
Substituting~\eqref{scaling} to~\eqref{eq:31-2}, we get
\begin{equation}
\det\left[
\psi_{j,j+1}(x_r^{(j)},x_s^{(j+1)})
\right]
=
\frac{W_T\left[\l^{(1)},\cdots, \l^{(M)}\right]}{Z'},
\end{equation}
where $Z'$ is the normalization constant. The weight $W_T$ has
the following form,
\begin{align}
  &\quad W_T
\left[\l^{(1)},\cdots,\l^{(M)}\right]\notag\\
&=\prod_{j=1}^M
\prod_{r=1}^{N}e^{-V_j(\l_r^{(j)})}
\prod_{j=1}^{M-1}\det[e^{c_j\l_r^{(j)}\l_s^{(j+1)}}]
\frac{\det[e^{c_0\e_r\l_s^{(1)}}]}{\Delta(\e)}\det
\left[e^{\frac{\sqrt{s_{M}}}{r_M\sqrt{T}}\l_r^{(M)}b_s}\right],
\label{WT}
\end{align}
where $\l^{(i)}$ means $\left\{\l^{(i)}_r\right\}_{r=1,\cdots,N}$ and  
\begin{equation}
  V_j(x)=\frac{s_j}{2}\left(\frac1{r_{j-1}}+\frac1{r_j}\right)x^2,
c_j=\frac{\sqrt{s_js_{j+1}}}{r_j}, c_0=\frac{\sqrt{s_1}}{2r_0},
\Delta(x)=\prod_{i<j}^{N}|x_i-x_j|.
\end{equation}
In~\eqref{WT}, we include the term $\Delta{(\e)}$ 
in order to get a finite expression when we consider the 
special case where ~\eqref{eq:4} in the later discussion.
Applying to the last term in $W_T$ the following formula,
\begin{equation}
  \lim_{\left\{z_i\right\}\rightarrow 1}
  \frac{\det\left[z_i^{\xi_j+N-j}\right]}{\det\left[z_i^{N-j}\right]}
=\prod_{1\le i<j\le N}\frac{\xi_i-\xi_j+j-i}{j-i},
\end{equation}
where $\{\xi_i\}\in \mathbb{C}^{\mathbb{N}}$,
one finds for large $T$,
\begin{align}
\det\left[e^{\frac{\sqrt{s_{M}}}{r_M\sqrt{T}}\l_r^{(M)}b_s}\right]
&=
\frac{\det\left[e^{\frac{\sqrt{s_{M}}}{r_M\sqrt{T}}\l_r^{(M)}b_s}\right]}
{\det\left[\left(e^{\frac{\l_i^{(M)}}{\sqrt{T}}}\right)^{N-j}\right]}
\det\left[\left(e^{\frac{\l_i^{(M)}}{\sqrt{T}}}\right)^{N-j}\right]\notag\\
&\sim\prod_{1\le i<j\le N}\frac{\sqrt{s_M}(b_i-b_j)}{j-i}
\frac{\Delta(\l^{(M)})}{T^{\frac{N(N-1)}{4}}}.
\end{align}
Thus we get
\begin{align}
  \lim_{T\rightarrow\i}\frac{W_T}{Z'}=
 \frac{1}{Z''}
\prod_{j=1}^M
\prod_{r=1}^{N}e^{-V_j(\l_r^{(j)})}
\prod_{j=1}^{M-1}\det[e^{c_j\l_r^{(j)}\l_s^{(j+1)}}]
\frac{\det[e^{c_0\e_r\l_s^{(1)}}]}{\Delta(\e)}
\Delta(\l^{(M)}).
\label{mm}
\end{align}
Up to a trivial constant, \eqref{mm} is equivalent to the weight 
of the eigenvalues of
the Hermitian multimatrix model,
\begin{equation}
  \prod_{j=1}^{M}e^{-\text{tr} V_j(H_j)}\prod_{j=1}^{M-1}
e^{\text{tr}c_jH_jH_{j+1}}
e^{\text{tr} c_0VH_1}dH_1\cdots dH_M.
\label{multi}
\end{equation}
Here $H_j (j=1, \cdots, M)$ is an $N\times N$ hermitian matrix
and $V=\text{diag}(\e)$.
This equivalence can be shown through
the Harish-Chandra,Itzykson-Zuber integral~\cite{HC,IZ},
\begin{equation}
  \int dU\exp\left(\text{tr}AUBU^{\dagger}\right)
\propto\frac{\det\left[\exp(a_ib_j)\right]}{\Delta(a)\Delta(b)},
\end{equation}
where $U$ is $N\times N$ unitary matrix, $dU$ is the Haar measure
and $A$ and $B$ are
$N\times N$ Hermitian matrices with eigenvalues
$a_i$ and $b_i$ respectively.
Note that in the case of one matrix model, \eqref{multi}
is the same form as the random matrix with deterministic 
source~\eqref{eq:1},~\eqref{eq:3} 
discussed in section \ref{PNGRMT}. 
We further transform the variables $s_j$ and 
$r_j$ as
\begin{equation}
  s_j=e^{2t_j},~~r_j=\frac{1-e^{2(t_j-t_{j+1})}}{2e^{-2t_{j+1}}},~~
  r_0=1/2,~~r_M\rightarrow\i,
\end{equation}
with $t_1=0$.
Then one finds \eqref{multi} can be represented as the form of
the Dyson's Brownian motion model,
\begin{equation}
  \prod_{j=1}^{M-1}\exp\left[\frac{-\text{tr}
\left\{H_{j+1}-e^{t_j-t_{j+1}}H_j\right\}^2}{1-e^{2(t_j-t_{j+1})}}\right]
\exp[-\text{tr}H_1^2+\text{tr}VH_1]dH_{1}\cdots dH_{M}.
\label{dbm}
\end{equation}
\subsection{Dynamical correlation function}
In the above discussion we construct the Dyson's Brownian motion model
with the deterministic source from the non-colliding Brownian motion
with the topology appropriate to the multi-layer PNG model with an 
external source.
Next we would like to analyze the process of the largest eigenvalue 
of this model since it corresponds to the equal time correlation of the 
height distribution in the PNG model.

The measure for the eigenvalues in~\eqref{dbm} is 
\begin{equation}
 \prod_{j=1}^{M-1} \det[\phi(t_{j},\l_r^{(j)};t_{j+1},
\l_s^{({j+1})})]_{r,s=1}^N
 e^{\sum_{j}{-\left(\l_j^{(1)}\right)^2}} 
\det\left[e^{\e_j \l_k^{(1)}}\right]_{j,k=1}^{N}
\det\left[\left(\l_j^{(M)}\right)^k\right]_{\begin{subarray}{c}
 j=1,...,N \\ k=0,...,N-1 \end{subarray}}
 ,
\label{edbm}
\end{equation}
where we denote the eigenvalue for the matrix $H_j$ as 
$\left\{\l_k^{(j)}\right\}_{k=1,\cdots,N}$
\begin{equation}
 \phi(t_i,x;t_j,y)
 =\begin{cases}
 \sqrt{\frac{e^{t_i-t_j}}{\pi(1-e^{2(t_i-t_j)})}}
 \exp\left[-\frac{(y-e^{t_i-t_j}x)^2}{1-e^{2(t_i-t_j)}}\right] 
& \text{for}~~{t_i\le t_j},\\
0&\text{for}~~{t_i> t_j}.
\end{cases}
\label{phi}
\end{equation}
Under this measure, we focus on the 
following probability,
\begin{equation}
  \P\left[\l_1^{(1)}\le s_1,\cdots ,\l_1^{(M)}\le s_M\right].
\label{core}
\end{equation}
We analyze this quantity by following the 
strategy of Johansson~\cite{Johansson2002p} although we 
can calculate it also by the method of Br{\'e}zin-Hikami~\cite{BH2}
with some effort.

In~\cite{Johansson2002p} it is shown that the correlation 
function~\eqref{core}
can be expressed as the Fredholm determinant for the case
where the measure is given in the determinantal form such as~\eqref{edbm}
by generalizing the method of Tracy-Widom~\cite{TW1998}. Applying the
procedure to~\eqref{edbm}, we get
\begin{equation}
 \P\left[\l_1^{(1)}\le s_1,\cdots ,\l_1^{(M)}\le s_M\right]=
\det\left[1+Kg\right], 
\end{equation}
where $g(t_j,x)=-\chi_{(s_j,\i)}(x)$. The kernel is
\begin{equation}
 K(t_r,x;t_s,y) = K'(t_r,x;t_s,y) -\phi(t_r,x;t_s,y),
\label{kernel}
\end{equation}
where
\begin{align} 
K'(t_r,x;t_s,y)
 &=
 \sum_{j,k=0}^{N-1} \Psi_j(x,t_r;t_M) (A^{-1})_{j,k} \Phi_k(t_0;t_s,y), 
\label{ker1}
\\
 \Psi_j(x,t_r;t_M) &= \int_{-\i}^{\i} \phi(t_r,x;t_M,y) y^j dy, 
\label{ker3}
\\
 \Phi_j(0;t_s,y) &= \int_{-\i}^{\i} e^{\e_{j+1}x} e^{-x^2}\phi(0,x;t_s,y) dx. 
\label{ker4}
\\
 A_{j,k} &= \int_{-\i}^{\i} dx dy e^{\e_{j+1}x} e^{-x^2} \phi(0,x;t_M,y) y^k, 
\label{ker2}
\end{align}
The analysis of the kernel using above equations seems to be difficult 
since~\eqref{ker1} involves the inverse matrix. However we can 
tackle this difficulty by improving the method of orthogonal
polynomial in RMT~\cite{Me1991}. 
We use efficiently the degree of freedom that
the value of a determinant is unchanged by elementary transformations.  
Let us define $F_{k,\e}(x)$ by
\begin{equation}
 F_{k,\e}(x) 
 = k!2^{k/2} \int_{\Gamma(\e')} \frac{dz}{2\pi i} 
   \frac{e^{-z^2/2+\sqrt{2}zx}}{\prod_{l=1}^{k+1}(z-\e_l')},
\label{F}
\end{equation}
where $\Gamma(\e')$ represents the contours enclosing all points
$\e_l'$ ($l=1,\cdots,k+1$) anticlockwise.
We also define $G_{k,\e}(x)$ by
\begin{equation}
 G_{k,\e}(x) 
 = \frac{2^{k/2}}{\sqrt{2\pi} i} e^{x^2}
   \int_{\g} dw e^{w^2/2-\sqrt{2}wx}\prod_{l=1}^k (w-\e_l'),
\label{G}
\end{equation}
where $\e_l'=\e_l/\sqrt{2}$ ($l=1,\cdots,N$) and 
$\g$ represents an arbitrary path running from $-i\i$ to $i\i$. 
The representations of $F_{j,\e}(x)$ and $G_{j,\e}(x)$ 
in~\eqref{F} and~\eqref{G} corresponds to the multiple
Hermite polynomial of type I and II respectively discussed
in~\cite{BK2} except some prefactors.
Note that when $\e_i=0$, both \eqref{F} and \eqref{G} become
the integral representation of the Hermite polynomial with degree $k$.

One can easily check that $F_k(x)$'s (resp. $G_k$'s) are 
linear combinations of $e^{\e_k x}$'s (resp. $x^k$'s).
Hence one finds
\begin{equation}
  \label{eq:32-2}
  \det[e^{\e_j\l_k^{(1)}}]_{j,k=1}^{N}
  =\text{const.}\times\det[F_{j,\e}(\l_k^{(1)})]_{\begin{subarray}{c}
 j=0,...,N-1 \\ k=1,...,N \end{subarray}},
\end{equation} 
\begin{equation}
  \label{eq:32-1}
  \det[(\l_j^{(M)})^k]_{\begin{subarray}{c}
 j=1,...,N \\ k=0,...,N-1 \end{subarray}}=\text{const.}\times 
\det[G_{k,e^{-t_M}\e}(\l_j^{(M)})]_{\begin{subarray}{c}
 j=1,...,N \\ k=0,...,N-1 \end{subarray}}.
\end{equation}
Applying the same procedure as above, one can show that the operator
$K$ on the right hand side in (\ref{core}) can be replaced by  
another operator $\tilde{K}$, with the kernel $\tilde{K}(t_r,x;t_s,y)$,
\begin{equation}
  \tilde{K}(t_r,x;t_s,y) = \tilde{K}'(t_r,x;t_s,y) -\phi(t_r,x;t_s,y),
\end{equation}
where
\begin{align}
  \tilde{K}'(t_r,x;t_s,y)
 &=
 \sum_{j,k=0}^{N-1} \tilde{\Psi}_j(x,t_r;t_M) (\tilde{A}^{-1})_{j,k} 
 \tilde{\Phi}_k(0;t_s,y),
\label{tK} 
\\
 \tilde{\Phi}_j(0;t_s,y) &= \int_{-\i}^{\i} 
F_{j,\e}(x) e^{-x^2}\phi(0,x;t_s,y) dx,
\\
  \tilde{\Psi}_j(x,t_r;t_M)&=\int_{-\i}^{\i}  
\phi(t_r,x;t_M,y)G_{j,e^{-t_M}\e}(y)dy,
\\
  \tilde{A}_{j,k}&=\int_{-\i}^{\i} dxdy F_{j,\e}(x) e^{-x^2}\phi(0,x;t_M,y)
G_{k,e^{-t_M }\e}(y).
\end{align} 
Substituting~\eqref{phi},~\eqref{F} and~\eqref{G} into these equations,
we get 
\begin{align}
\label{tA}
\tilde{\Phi}_j(0;t_s,y)&=e^{-(j+1/2)t_s} e^{-y^2} F_{j,e^{-t_s}\e}(y),\\
\tilde{\Psi}_j(x,t_r;t_M)&=e^{-(j+1/2)(t_M-t_r)} G_{j,e^{-t_r}\e}(x)
\end{align}
and
\begin{equation}
 \tilde{A}_{j,k}
 =
 e^{-(j+1/2)t_M} \int_{-\i}^{\i} dx~F_{j,\e}(x) G_{k,\e}(x)e^{-x^2}  
 =
 e^{-(j+1/2)t_M} \sqrt{\pi} 2^j j! \delta_{j,k}.  
\label{tC}
\end{equation}
We chose $F$ and $G$ in such a way that $\tilde{A}_{j,k}$
becomes diagonal and can hence be easily invertible.
Note that when $\e_i=0$ and $t_M=0$, 
~\eqref{tC} represents the orthogonality of the Hermite polynomials. 
Our treatment here generalizes the method 
of orthogonal polynomials for the multimatrix model between 
GUE~\cite{EM1999,FNH1999} to the case with an external source. 

Substituting~\eqref{tA}--~\eqref{tC} to~\eqref{tK}, one gets 
\begin{align}
 \tilde{K}'(t_r,x;t_s,y)
 &=
 \sum_{j=0}^{N-1} \frac{e^{(j+1/2)t_M}}{\sqrt{\pi}2^j j!} 
                  \tilde{\Psi}_j(x,t_r;t_M) \tilde{\Phi}_j(0;t_s,y) \notag\\
 &=
 e^{-y^2} \sum_{j=0}^{N-1} G_{j,e^{-t_r}\e}(x) F_{j,e^{-t_s}\e}(y) 
                           e^{-(j+1/2)(t_s-t_r)} \notag\\
 &=
 e^{x^2-y^2} \frac{\sqrt{2}e^{\frac12 (t_r+t_s)}}{(2\pi i)^2}
 \int_{\Gamma(e^{-t_s}\e')} dz \int_{\g} dw
 e^{w^2/2-\sqrt{2}wx-z^2/2+\sqrt{2}zy}\sum_{j=0}^{N-1}\frac{\prod_{l=1}^{j}
e^{t_r}w-\e'_l}{\prod_{l=1}^{j+1}e^{t_s}z-\e'_l},
\label{tKK}
\end{align}
where in the last equality we use~\eqref{F} and~\eqref{G}. 
The summation term in the integrand can be deformed to
\begin{align}
&~~\sum_{j=0}^{N-1}\frac{\prod_{l=1}^{j}
e^{t_r}w-\e'_l}{\prod_{l=1}^{j+1}e^{t_s}z-\e'_l}\notag\\
&=\prod_{l=1}^{N}\left(\frac{e^{t_r}w-\e'_l}{e^{t_s}z-\e'_l}\right)
\left\{
\frac1{e^{t_r}w-\e'_N}+\frac{e^{t_s}z-\e'_N}{(e^{t_r}w-\e'_{N-1})
(e^{t_r}w-\e'_{N})}+\cdots+
\frac{(e^{t_s}z-\e'_2)\cdots(e^{t_s}z-\e'_N)}
{(e^{t_r}w-\e'_1)\cdots(e^{t_r}w-\e'_N)}
\right\} \notag\\
&=
\prod_{l=1}^{N}\left(\frac{e^{t_r}w-\e'_l}{e^{t_s}z-\e'_l}\right)
\left\{
\frac1{e^{t_r}w-e^{t_s}z}-\prod_{l=1}^{N}
\left(\frac{e^{t_s}z-\e'_l}{e^{t_r}w-\e'_l}\right)
\frac1{e^{t_r}w-e^{t_s}z}
\right\}.
\label{sp}
\end{align}
From~\eqref{tKK} and~\eqref{sp} one finally finds the 
double integral formula of the kernel,
\begin{align}
  \label{eq:32-4}
 &\quad\tilde{K}'(t_r,x;t_s,y)\notag\\
&=
e^{x^2-y^2} \frac{\sqrt{2}e^{\frac12 (t_r-t_s)}}{(2\pi i)^2}
 \int_{\Gamma(e^{-t_s}\e')} dz \int_{\g} dw
 \prod_{l=1}^N \frac{e^{t_r}w-\e_l/{\sqrt{2}}}{e^{t_s}z-\e_l/{\sqrt{2}}} 
 \frac{1}{w e^{t_r-t_s}-z}
 e^{w^2/2-\sqrt{2}wx-z^2/2+\sqrt{2}zy}, 
\end{align}
where we use the fact that the second term in~\eqref{sp} does not 
contribute to the double integral in~\eqref{tKK}. In the one matrix case
where $t_r=t_s$, the kernel in~\eqref{eq:32-4} becomes equivalent to 
the kernel~\eqref{eq:6} obtained by Br\'ezin-Hikami  
except the prefactor irrelevant for the determinant.

\subsection{GOE$^2$/GUE transition}
\label{12transition}
We consider the asymptotic form of the kernel~\eqref{kernel} under the 
edge scaling.
Hereafter we consider the case where $\e_1=\e, \e_j=0$ ($j=2,...,N$).
Rescaling the integration variables $w,z$, the kernel~\eqref{eq:32-4} is 
rewritten as
\begin{align}
 \tilde{K}'(t_1,x;t_2,y)
 &=
 e^{x^2-y^2} \frac{\sqrt{2N}e^{\frac12 (t_1-t_2)}}{(2\pi i)^2}
 e^{(N-1)(t_1-t_2)} \notag\\ 
&\quad \times\int_{\Gamma'_N}dz\int_{\g}dw\frac{z}{w}
\frac{e^{t_1} w -\e/\sqrt{2N}}{e^{t_2}z-\e/\sqrt{2N}} 
 \frac{1}{w e^{t_1-t_2}-z}
\frac{e^{\sqrt{2N}zy-2Nz}}{e^{\sqrt{2N}wx-2Nw}}
e^{N\left\{f(w)-f(z)\right\}}, 
\label{331}
\end{align}
where the contour $\Gamma'_N$ encloses the origin and 
$e^{-t_2}\frac{\e'}{\sqrt{N}}$ anticlockwise and
\begin{equation}
  f(w)\equiv \frac{w^2}{2}-2w+\ln w.
\end{equation}
We analyze the asymptotic behavior of~\eqref{331} by applying 
the saddle point method. We scale the variables in~\eqref{331}
as 
\begin{gather}
 x = \sqrt{2N}+ \frac{\xi_1}{\sqrt{2}N^{1/6}}, 
 \quad 
 y = \sqrt{2N}+ \frac{\xi_2}{\sqrt{2}N^{1/6}}, \notag\\ 
 t_1 = \frac{\t_1}{N^{1/3}}, \quad t_2 = \frac{\t_2}{N^{1/3}}, \notag\\
 \e = \sqrt{2N} \left(1-\frac{\omega}{N^{1/3}}\right).
\label{esc}
\end{gather}
The critical point of the function $f(w)$,    
\begin{equation}
  w_c=1,
\end{equation}
turns out to be the double critical point,
\begin{equation}
  f'(w_c)=f''(w_c)=0.
\end{equation}
Thus by deforming the paths of the integration in~\eqref{331} in a way 
that they cross the critical point,
\begin{align}
 &w=1-\frac{iw_1}{N^{\frac13}},\\
 &z=1+\frac{iw_2}{N^{\frac13}},
\end{align}
we get
\begin{equation}
  e^{N\left\{f(w)-f(z)\right\}}\sim e^{\frac{i}{3}(w_1^3+w_2^3)},
\end{equation}
where we use
\begin{align}
  f(w)&\sim f(w_c)+f'(w_c)(w-w_c)+\frac{f''(w_c)}{2!}(w-w_c)^2+
\frac{f'''(w_c)}{3!}(w-w_c)^3, \notag\\
    &=\frac{-3}{2}+\frac{i}{3N}w_1^3.
\end{align}
For other terms in~\eqref{331} we obtain the asymptotic forms under 
the scaling~\eqref{esc},
\begin{equation}
\frac{e^{t_1}w-\e/\sqrt{2N}}{e^{t_2}z-\e/{\sqrt{2N}}}\sim 
\frac{\t_1+\o-iw_1}{\t_2+\o+iw_2},  
\end{equation}
\begin{equation}
  \frac{1}{we^{t_1-t_2}-z}\sim -\frac{N^{\frac13}}{\t_2-\t_1+iw_1+iw_2},
\end{equation}
\begin{equation}
  \frac{e^{\sqrt{2N}zy-2Nz}}{e^{\sqrt{2N}wx-2Nw}}\sim
e^{N^{\frac13}(\xi_2-\xi_1)+i\xi_1w_1+i\xi_2w_2},
\end{equation}
\begin{equation}
  e^{x^2-y^2+\frac12(t_1-t_2)+(N-1)(t_1-t_2)}\sim
e^{2N^{\frac13}(\xi_1-\xi_2)+N^{\frac23}(\t_1-\t_2)}.
\end{equation}
Hence we eventually finds
\begin{align}
 \tilde{K}' 
 &\sim 
 \sqrt{2}N^{1/6} e^{N^{2/3}(\t_1-\t_2)+N^{1/3}(\xi_1-\xi_2)} \notag\\
 &\quad \times
 \int_{-\i}^{\i} \frac{dw_1}{2\pi} 
 \int_{-\i}^{\i} \frac{dw_2}{2\pi}
 \left(-\frac{1}{\t_2-\t_1+i(w_1+w_2)}+\frac{1}{\omega+\t_2+iw_2}\right)
 e^{i(\xi_1 w_1+\xi_2 w_2)+\frac{i}{3}(w_1^3+w_2^3)} \notag\\
 &=\sqrt{2}N^{1/6} e^{N^{2/3}(\t_1-\t_2)+N^{1/3}(\xi_1-\xi_2)} \notag\\
 &\quad \times
 \int_0^{\i}d\l e^{-\l(\t_1-\t_2)}\Ai(\xi_1+\l)\Ai(\xi_2+\l)
 +\Ai(\xi_1)\int_0^{\i}d\l e^{-(\o+\t_2)\l}\Ai(\xi_2-\l),
\label{agoe2}
\end{align}
where we use the integral representation of the Airy function,
\begin{equation}
  \Ai(x)=\int_{-\i}^{\i}d\l e^{i\l x+\frac{i}{3}\l^3}.
\end{equation}

It is also necessary to consider the asymptotics of $\phi(t_1,x;t_2,y)$
in~\eqref{phi}
under the same scaling as in~\eqref{esc}. One has
\begin{align}
 -\frac{1}{1-e^{2(t_1-t_2)}}
 &\sim 
 -\frac{N^{1/3}}{2(\t_2-\t_1)}
 \left\{1+\frac{1}{N^{1/3}}(\t_2-\t_1)
          +\frac{1}{3N^{2/3}}(\t_2-\t_1)^2+...\right\},\notag \\
 (e^{t_1-t_2}x-y)^2
 &\sim
 2N^{1/3} \Bigl\{ (\t_2-\t_1)^2-\frac{1}{N^{1/3}}(\t_2-t_1)^3 \notag\\
 &\quad +\frac{1}{N^{1/3}}(\xi_2-\xi_1)(\t_2-\t_1)
 -\frac{1}{2 N^{2/3}}(\xi_2-\xi_1)(\t_2-\t_1)^2
 +\frac{7}{12 N^{2/3}}(\t_2-\t_1)^4 \notag\\
 &\quad +\frac{1}{4N^{2/3}}(\xi_2-\xi_1)^2
 +\frac{\xi_1}{N^{2/3}}(\t_2-\t_1)^2 \Bigr\}.
\end{align}
Substituting these into (\ref{phi}), one gets
\begin{align}
\phi 
=
\begin{cases}
\sqrt{2}N^{1/6} e^{N^{2/3}(\t_1-\t_2)+N^{1/3}(\xi_1-\xi_2)}
\int_{-\i}^{\i}d\l e^{-\l(\t_1-\t_2)}\Ai(\xi_1+\l)\Ai(\xi_2+\l)&
{\text{for}}~~\t_1\le\t_2.\\
0 & {\text{for}}~~\t_1>\t_2 .
\end{cases}
\end{align}
Thus we finally find 
\begin{equation}
 \sqrt{2}N^{1/6} \tilde{K}
 \to
 e^{N^{2/3}(\t_1-\t_2)+N^{1/3}(\xi_1-\xi_2)}
 \K(\t_1,\xi_1;\t_2,\xi_2)
\end{equation}
where $ \K(\t_1,\xi_1;\t_2,\xi_2)$ is the same as the one appearing 
in the analysis of the PNG model with external source~\eqref{Kernel12}.
Since the prefactor is irrelevant to the value of Fredholm 
determinants, we have shown that the fluctuation of the PNG height 
described by (\ref{detKlim_o0})-(\ref{K2def}) is equivalent 
to the fluctuation of the largest eigenvalue of the 
multi-matrix model (\ref{dbm}) with an appropriate choice of $V$.

\section{Conclusion}
We have presented a random matrix model with a deterministic source,
the largest eigenvalue of which describes the height fluctuations 
of the PNG model with an external source.
Depending on the value of the deterministic source, the distribution of 
the largest eigenvalue in this model becomes the GUE Tracy-Widom
distribution and the GOE$^2$. It also describes the transition 
between these two,  which could not be understood by the previous 
interpretation of the GOE$^2$ as ``the square of GOE''.
Our model gives not only another representation of the GOE$^2$ 
but also a unified picture including GUE, GOE$^2$ and 
their transition.  

We have also considered a vicious walk model which has a similar topology 
to that of the multi-layer PNG model. In an appropriate limit, 
this leads us to a Dyson's Brownian motion model with a deterministic
source as an initial condition. 
At last we have found the process of the largest eigenvalue in the edge 
scaling of this model describes the multi-point joint distributions 
of the PNG model with an external source. 

\section*{Acknowledgment}
The authors would like to thank M. Katori, T. Nagao, A. R{\'a}kos, 
G. M. Sch{\"u}tz and H. Spohn for fruitful discussions and comments.  

The work of T.I. is partly supported by the Grant-in-Aid for JSPS
Fellows, the Ministry of Education, Culture, Sports, Science and 
Technology, Japan.
The work of T.S. is partly supported by the Grant-in-Aid for Young 
Scientists (B), the Ministry of Education, Culture, Sports, Science and 
Technology, Japan.

\newpage
%%%%%%%%%%%%%%%%%%%%%%%%%%%%%%%%%
%%% Figure Captions           %%%
%%%%%%%%%%%%%%%%%%%%%%%%%%%%%%%%%
\begin{large}
\noindent
Figure Captions
\end{large}
%%% Fig. 1 %%%%%%%%%%%%%%%%%%%%%%

\vspace{10mm}
\noindent
Fig.1: Rules of the discrete PNG model. Here we draw the solid lines
using $h([r],t)$.

\vspace{1cm}
\noindent
Fig.2: Typical situation in the PNG droplet with an external source.
In this model the rate of growth at the left edge($r=-t+1$) is
higher than that at other places since the parameter of a nucleation 
at the left edge, $\a\sqrt{q}$, is larger than that at other
points, $\sqrt{q}$.

\vspace{1cm}
\noindent
Fig.3: 
\noindent
(a) Rule of the multi-layer PNG model. Due to the rule(c) of the PNG model,
the part illustrated by the thick lines vanishes in the first layer.
This part, however, is recovered by the nucleation of the layer below.
For the second and subsequent layers, we follow the same procedure. 

\noindent
(b) A typical configuration of the PNG droplet with an external 
source for $\a=1, q=\frac14$ and $t=200$.

\vspace{1cm}
\noindent
Fig.4: Non-colliding Brownian motion in the scaling limit~\eqref{scaling}.
%%%%%%%%%%%%%%%%%%%%%%%%%%%%%%%%%
%%% Figures                   %%%
%%%%%%%%%%%%%%%%%%%%%%%%%%%%%%%%%
%%%%%%%%%%% Fig. 1 Rule of PNG model %%%%%%%%%
\newpage
\renewcommand{\thepage}{Figure 1}
\begin{picture}(400,600)
\psfrag{k}{$k$}
\put(0,400){\resizebox{14cm}{7cm}{\includegraphics{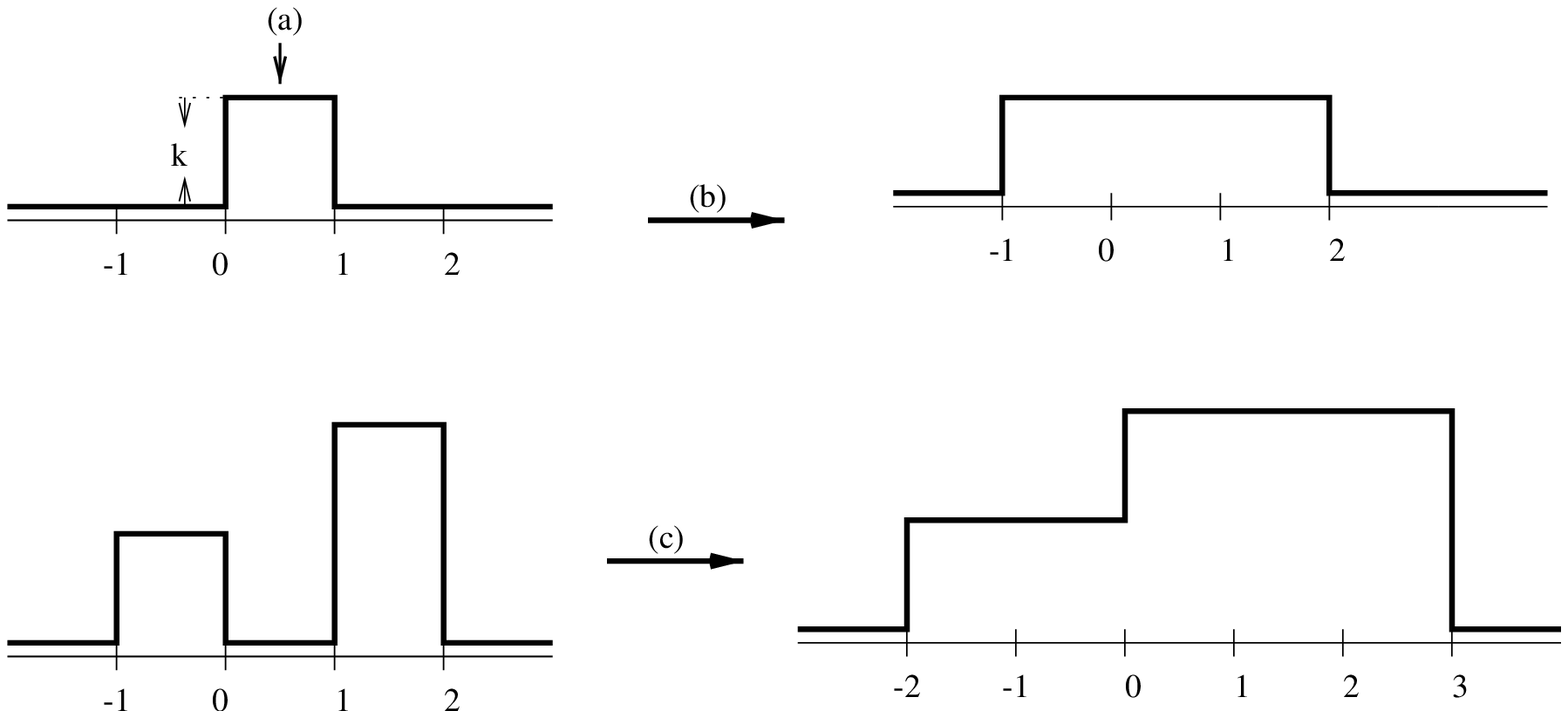}}}
\end{picture}
\newpage
\renewcommand{\thepage}{Figure 2}
\begin{picture}(400,600)
\psfrag{k}{$\a\sqrt{q}$}
\psfrag{q}{$\sqrt{q}$}
\psfrag{x}{$\times$}
\psfrag{-t+1}{$-t+1$}
\psfrag{t-1}{$t-1$}
\put(0,400){\resizebox{14cm}{7cm}{\includegraphics{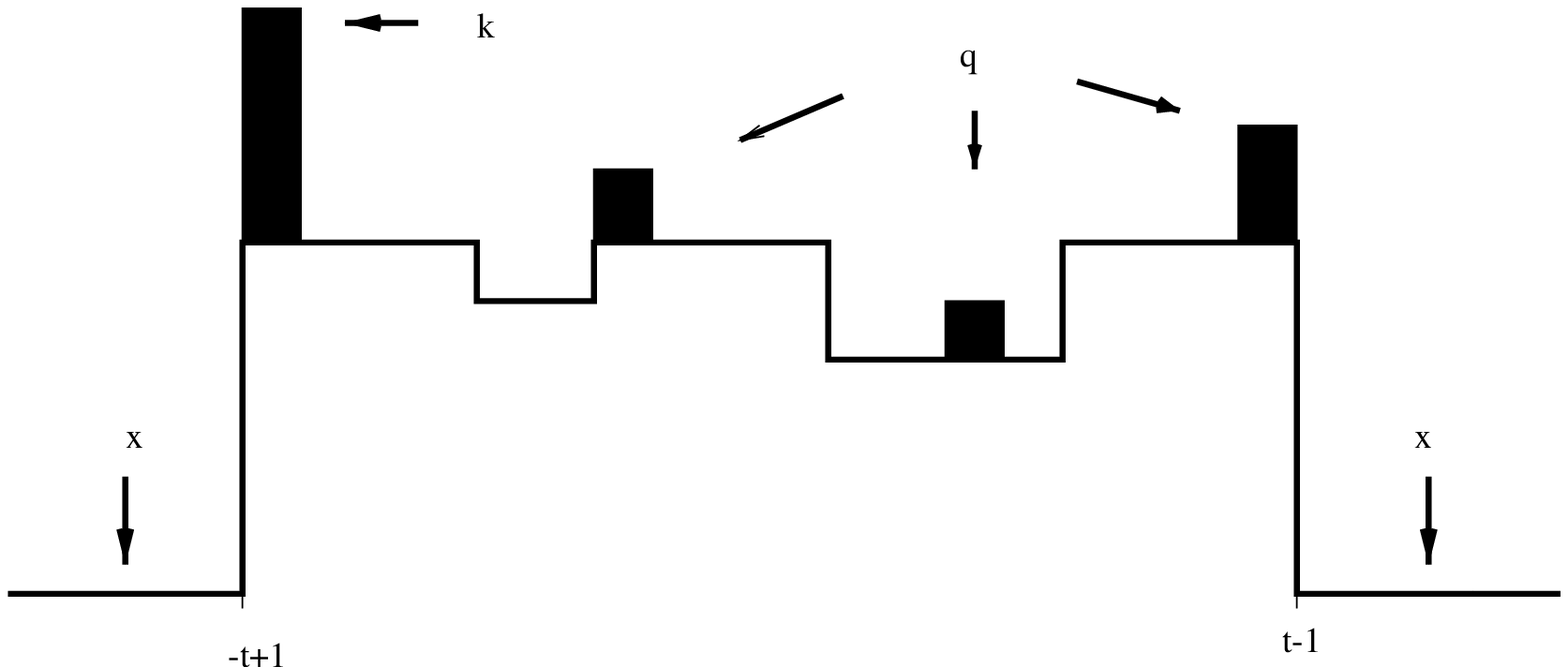}}}
\end{picture}
\newpage
\renewcommand{\thepage}{Figure 3}
\begin{picture}(400,600)
\put(0,580){(a)}
\put(0,500){\resizebox{14cm}{3cm}{\includegraphics{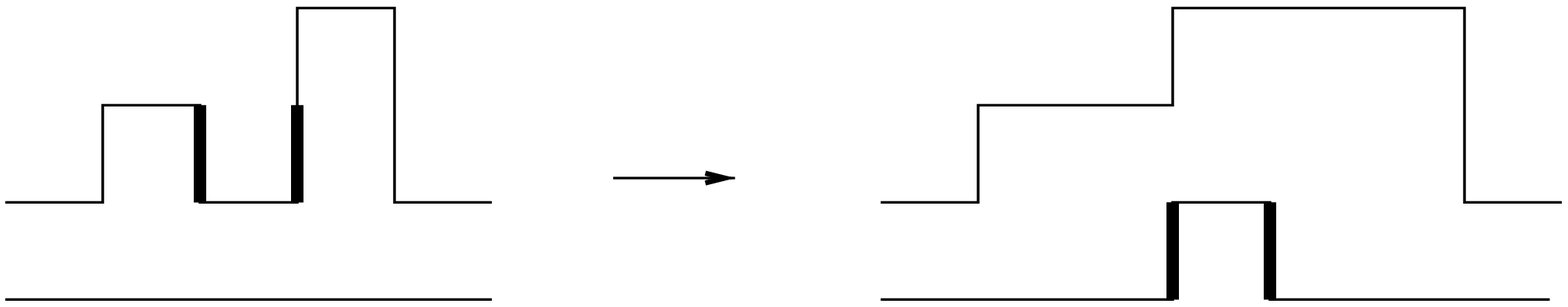}}}
\put(0,400){(b)}
\put(0,100){\resizebox{14cm}{10cm}{\includegraphics{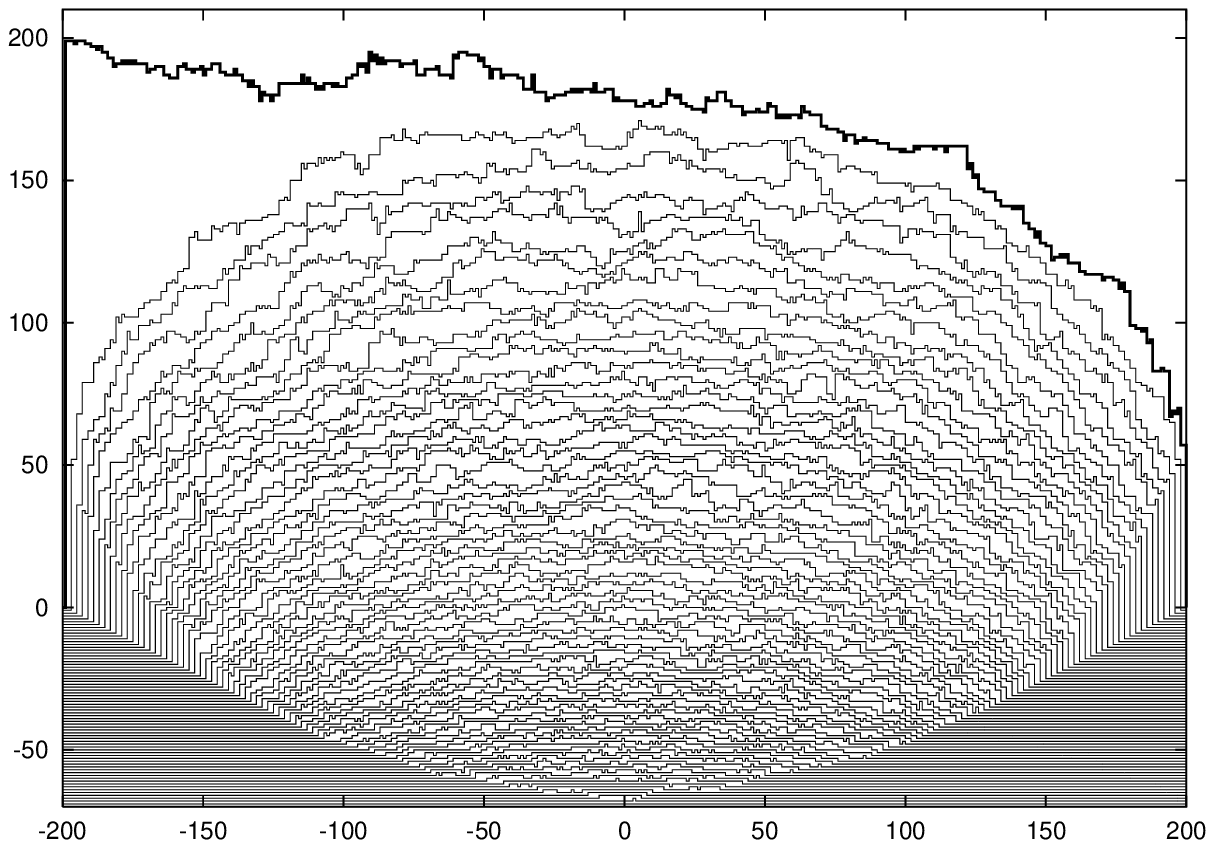}}}
%\put(0,10){\includegraphics{fig2.eps}}
\end{picture}
\newpage
\renewcommand{\thepage}{Figure 4}
\begin{picture}(400,600)
\psfrag{a1T}{$\frac{\sqrt{T}\e_1}{2}$}
\psfrag{a2T}{$\frac{\sqrt{T}\e_2}{2}$}
\psfrag{aN-1T}{$\frac{\sqrt{T}\e_{N-1}}{2}$}
\psfrag{aNT}{$\frac{\sqrt{T}\e_N}{2}$}
\psfrag{0}{$0$}
\psfrag{1}{$1$}
\psfrag{2}{$2$}
\psfrag{M}{$M$}
\psfrag{M+1}{$M+1$}
\psfrag{x}{$x$}
\psfrag{b1}{$b_1$}
\psfrag{b2}{$b_2$}
\psfrag{b3}{$b_{N-1}$}
\psfrag{b4}{$b_{N}$}
\psfrag{t1T}{$r_1T$}
\psfrag{xx}{$\propto\sqrt{T}$}
\psfrag{t}{$t$}
\put(0,300){\resizebox{14cm}{9cm}{\includegraphics{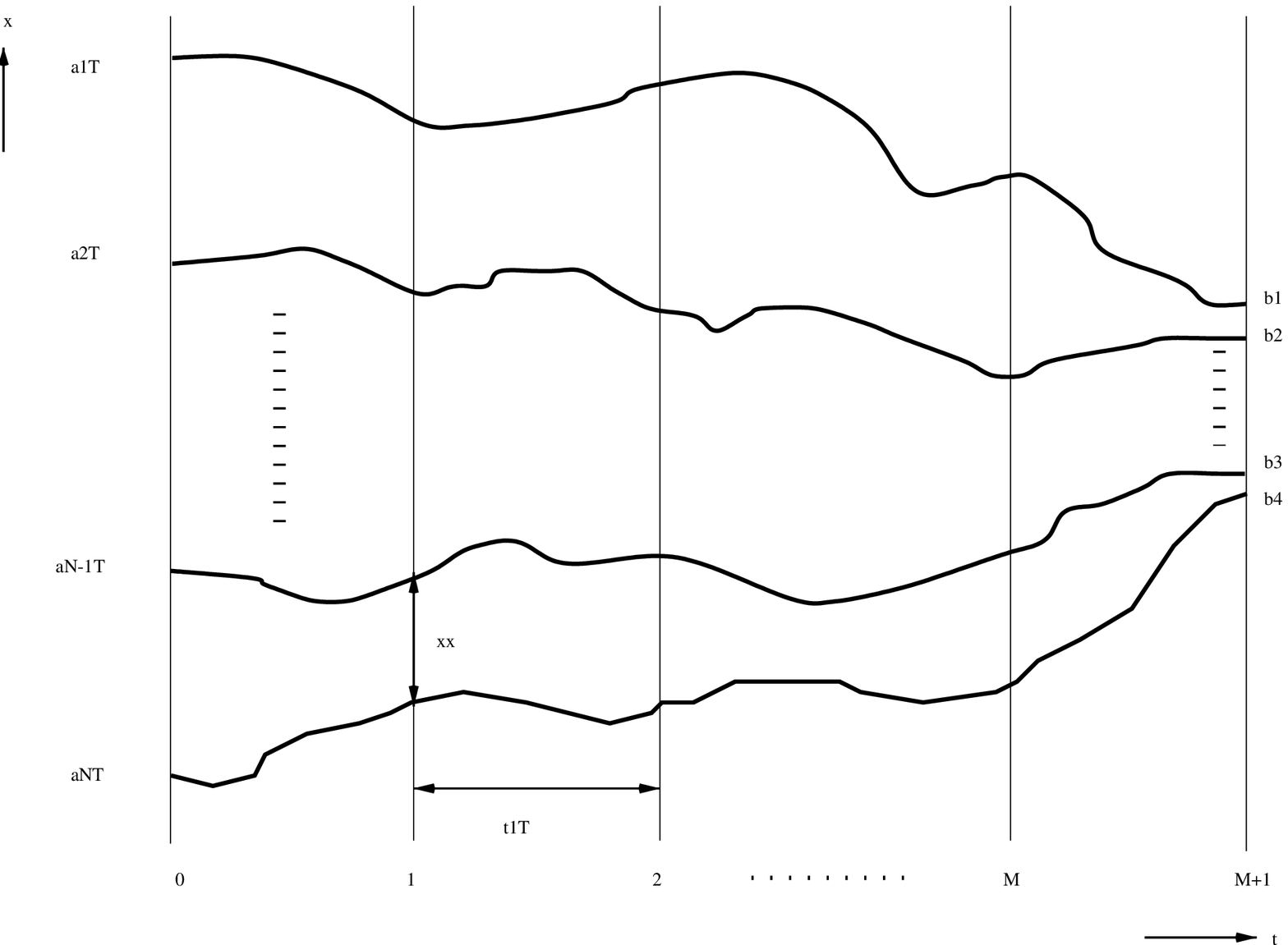}}}
\end{picture}


\begin{thebibliography}{10}
\bibitem{KPZ1986}
{M. Kardar, G. Parisi and Y. C. Zhang}.
\newblock Dynamic scaling of growing interfaces.
\newblock {\em Phys. Rev. Lett.}, 56:889--892, 1986.

\bibitem{PS2000b}
M.~Pr{\"a}hofer and H.~Spohn.
\newblock Statistical self-similarity of one-dimensional growth processes.
\newblock {\em Physica A}, 279:342--352, 2000.

\bibitem{Jo2000}
K.~Johansson.
\newblock Shape fluctuations and random matrices.
\newblock {\em Commun. Math. Phys.}, 209:437--476, 2000.

\bibitem{BDJ1999}
{J. Baik, P. A. Deift and K. Johansson}.
\newblock On the distribution of the length of the longest increasing
  subsequence in a random permutation.
\newblock {\em J. Amer. Math. Soc.}, 12:1119--1178, 1999.

\bibitem{BR2001a}
J.~Baik and E.~M. Rains.
\newblock Algebraic aspects of increasing subsequences.
\newblock {\em Duke Math. J.}, 109:1--65, 2001.

\bibitem{BR2001b}
J.~Baik and E.~M. Rains.
\newblock The asymptotics of monotone subsequences of involutions.
\newblock {\em Duke Math. J.}, 109:205--281, 2001.

\bibitem{BR2001c}
J.~Baik and E.~M. Rains.
\newblock Symmetrized random permutations.
\newblock In P.~M. Bleher and A.~R. Its, editors, {\em Random Matrix Models and
  Their Applications}, pages 1--29, 2001.

\bibitem{TW1994}
C.~A. Tracy and H.~Widom.
\newblock Level-spacing distributions and the {Airy} kernel.
\newblock {\em Commun. Math. Phys.}, 159:151--174, 1994.

\bibitem{TW1996}
C.~A. Tracy and H.~Widom.
\newblock On orthogonal and symplectic matrix ensembles.
\newblock {\em Commun. Math. Phys.}, 177:727--754, 1996.

\bibitem{PS2000a}
M.~Pr{\"a}hofer and H.~Spohn.
\newblock Universal distributions for growth processes in 1+1 dimensions and
  random matrices.
\newblock {\em Phys. Rev. Lett}, 84:4882--4885, 2000.

\bibitem{PS2002b}
M.~Pr{\"a}hofer and H.~Spohn.
\newblock Scale invariance of the {PNG} droplet and the {A}iry process.
\newblock {\em J. Stat. Phys.}, 108:1071--1106, 2002.

\bibitem{Johansson2002p}
K.~Johansson.
\newblock Discrete polynuclear growth and determinantal processes.
\newblock {\em Com. Math. Phys.}, 242: 277--329, 2003.

\bibitem{SI2004}
T.~Sasamoto and T.~Imamura.
\newblock Fluctuations of the One-Dimensional Polynuclear Growth Model 
in Half-Space
\newblock {\em J. Stat. Phys.}, 115:749-803, 2004. 

\bibitem{BR2000}
J.~Baik and E.~M. Rains.
\newblock Limiting distributions for a polynuclear growth model with external
  sources.
\newblock {\em J. Stat. Phys}, 100:523--541, 2000.

\bibitem{PS2002a}
M.~Pr{\"a}hofer and H.~Spohn.
\newblock Current fluctuations for the totally asymmetric simple exclusion
  process.
\newblock In V.~Sidoravicius, editor, {\em In and out of equilibrium, vol. 51
  of \it Progress in Probability}, pages 185--204, 2002.

\bibitem{NS2004p}
T.~Nagao and T.~Sasamoto.
\newblock Asymmetric simple exclusion process and modified random
matrix ensembles.
\newblock{Nucl. Phys. B}, 699: 487--502, 2004.

\bibitem{Forrester2000p}
{P. J. Forrester}.
\newblock Painlev\'e transcendent evaluation of the scaled distribution
of the smallest eigenvalue in the Laguerre orthogonal and symplectic 
ensembles.
nlin.SI/0005064.

\bibitem{BH1}
{E. Br{\'e}zin, S. Hikami and A. Zee.}
\newblock{ Universal correlations for deterministic plus random Hamiltonians.}
\newblock{\em Phys. Rev. E}, 51: 5442-5452, 1995.

\bibitem{BH2}
{E. Br{\'e}zin and  S. Hikami.}
\newblock{Spectral form factor in a random matrix theory.}
\newblock{\em Phys. Rev. E}, 55: 4067-4083, 1997.

\bibitem{BH3}
{E. Br{\'e}zin and  S. Hikami.}
\newblock{Extension of level-spacing universality.}
\newblock{\em Phys. Rev. E}, 56: 264-269, 1997.

\bibitem{BH4}
{E. Br{\'e}zin and S. Hikami.}
\newblock{Universal singularity at the closure of 
a gap in a random matrix theory.}
\newblock{\em Phys. Rev. E}, 57: 4140-4149, 1998.

\bibitem{BH5}
{E. Br{\'e}zin and S. Hikami.}
\newblock{Level spacing of random matrices in an external source.}
\newblock{\em Phys. Rev. E}, 58: 7176-7185, 1998.

\bibitem{BH6}
{E. Br{\'e}zin, S. Hikami and A. Zee.}
\newblock{Oscillating density of states near zero energy 
for matrices made of blocks with possible application to 
the random flux problem.}
\newblock{\em Nuc. Phys. B}, 464: 411-448, 1996.

\bibitem{BH7}
{E. Br{\'e}zin and S. Hikami.}
\newblock{Correlations of nearby levels induced by a random potential.}
\newblock{\em Nuc. Phys. B}, 479: 697-706, 1996.

\bibitem{PZ1}
{P. Zinn-Justin.}
\newblock{Random Hermitian matrices in an external field.}
\newblock{\em Nuc. Phys. B}, 497: 725-732,  1997. 


\bibitem{PZ2}
{P. Zinn-Justin.}
\newblock{Universality of correlation functions of hermitian 
random matrices in an external field .}
\newblock{\em Com. Math. Phys.} 194: 631-650, 1998.

\bibitem{BK1}
{P. M. Bleher and A. B. J. Kuijlaars.}
\newblock{Random matrices with external source and multiple 
orthogonal polynomials.}
\newblock{\em Int. Math. Res. Not.} 109-129, 2004.

\bibitem{BK2}
{P. M. Bleher and A. B. J. Kuijlaars.}
\newblock{Integral representations for multiple Hermite and 
multiple Laguerre polynomials.}
math.CA/0406616.

\bibitem{BK3}
{P. M. Bleher and A. B. J. Kuijlaars.}
\newblock{Large $n$ limit of Gaussian random matrices with 
external source, part I.}
math-ph/0402042.

\bibitem{BK4}
{A. I. Aptekarev, P. M. Bleher and A. B. J. Kuijlaars.}
\newblock{Large $n$ limit of Gaussian random matrices with 
external source, part II.}
math-ph/0408041.

\bibitem{Forrester1993}
{P. J. Forrester}.
\newblock{The spectrum  edge of random-matrix ensembles.}
\newblock{\em Nuc. Phys. B},402: 709-728, 1994.

\bibitem{BBAP2004p}
{J. Baik, G. Ben Arous, S. Peche}
\newblock{Phase transition of the largest eigenvalue for 
non-null complex sample covariance matrices.}
math.PR/0403022.

\bibitem{IS2004p}
T. Imamura and T. Sasamoto.
\newblock{Fluctuations of the one-dimensional polynuclear growth model 
with external sources.}
\newblock{Nucl. Phys. B}, 699: 503--544, 2004.
%math-phys/0406001.

\bibitem{Mac1994}
A.~M.~S. Mac\^edo.
\newblock Universal parametric correlations at the soft edge of spectrum of
  random matrix ensembles.
\newblock {\em Europhys. Lett.}, 26:641--646, 1994.

\bibitem{KT1}
M. Katori and H. Tanemura.
\newblock{Scaling limit of vicious walks and two-matrix model.}
\newblock{\em Phys. Rev. E}, 66: Art. No. 011105 Part1, 2002.

\bibitem{KT2}
M. Katori and H. Tanemura.
\newblock{Symmetry of matrix-valued stochastic processes and 
noncolliding diffusion particle systems.}
\newblock{\em J. Math. Phys.}, 45: 3058-3085, 2004.

\bibitem{HC}
Harish-Chandra.
\newblock{Differential operators on a semisimple Lie algebra.}
\newblock{\em Amer. J. Math.}, 79: 87-120, 1957.

\bibitem{IZ}
C. Itzykson and J.B. Zuber.
\newblock Planar approximation II. 
\newblock{\em J. Math. Phys.},  21: 411-421, 1980.
\bibitem{TW1998}
C.~A. Tracy and H.~Widom.
\newblock Correlation functions, cluster functions, and spacing distributions
  for random matrices.
\newblock {\em J. Stat. Phys.}, 92:809--835, 1998.

\bibitem{Me1991}
M.~L. Mehta.
\newblock {\em Random Matrices}.
\newblock Academic, 2nd edition, 1991.

\bibitem{EM1999}
B. Eynard and M. L. Mehta.
\newblock{Matrices coupled in a chain: I. Eigenvalue correlations.}
\newblock{\em J. Phys. A}, 31: 4449-4456, 1998.

\bibitem{FNH1999}
{P. J. Forrester, T. Nagao and G. Honner}.
\newblock Correlations for the orthogonal-unitary and symplectic 
transitions at the hard and soft edges.
\newblock {\em Nucl. Phys. B}, 553:601--643, 1999.
\end{thebibliography}
\end{document}